\documentclass[onecolumn,12pt]{article}

\usepackage{graphicx}
\usepackage{bm}
\usepackage{amsmath}
\usepackage{amssymb}
\usepackage{float}
\usepackage{caption}
\usepackage{subcaption}
\usepackage{cite}

\usepackage{authblk}
\title{More than one Author with different Affiliations}
\author[1,2]{Rukmani Bai\thanks{rukmani@prl.res.in}}
\author[1,2]{Pankaj Bhalla\thanks{pankaj@prl.res.in}}
\author[1]{Navinder Singh\thanks{navinder@prl.res.in}}
\affil[1]{\it Theoretical Physics Division, Physical Research Laboratory, Ahmedabad-380009, India}
\affil[2]{\it Indian Institute of Technology Gandhinagar-382424, India}

\begin{document}
	
\title{\Large Theory of the electron phonon relaxation time in cuprates: Reproducing the observed temperature behaviour}
\maketitle

\begin{abstract}
We have studied the temperature dependence of the  rate of energy transfer from electronic sub-system to phononic sub-system in the case of cuprates, when the system is photo-excited by a femtosecond laser pulse. In the pseudogap state, taking the electronic dispersion {\it as linear} near the nodal points of the Brillouin zone, we show that the rate of energy transfer from electronic sub-system to phononic sub-system is proportional to $T^{5}$ at lower temperatures ($T<<T_0$), and is proportional to $T$ at higher temperatures ($T>>T_0$), here $T_0$ is the Debye temperature for cuprates. The linear electronic dispersion in the pseudogap state introduces new terms in the expression of energy transfer as given by M. I. KAGANOV et.al. \cite{kaganov}. {\it But the leading terms are the same which are found in the case of metals in the above reference.} The electron-phonon relaxation time follows $T^{-3}$ law for cuprates which agrees well with the experimental results \cite{demsar,Jdemsar}.
\end{abstract}
\section{Introduction}
Pump probe spectroscopy is a useful tool to study the electronic dynamics of a material \cite{node,ttm,fs,fs0,fs1,fs2,fs3,fs4,fs5,fs6,fs7,fs8}. In this spectroscopy, first an ultrafast femtosecond laser pulse is divided into two ultrashort pulses, one is  pump pulse and another is probe pulse, with a tunable time delay between them. Then these pulses are injected into a sample. The pump pulse excites the quasiparticles in the material and the probe pulse gives information of the subsequent quasiparticles dynamics. The excitation by the pump pulse brings the system into non-equilibrium state which after a time scale of picoseconds, comes into equilibrium through the relaxation process. During excitation process, electron temperature rises above the equilibrium phonon temperature. This happens because electrons have smaller heat capacity as compared to the phonons \cite{ttm}. The excited electrons (or electronic subsystem) relax first within themselves and then they relax with phononic degrees-of-freedom.\\\\
The pump probe spectroscopy is being used to understand the non-equilibrium transport properties of high temperature superconductors like cuprates \cite{node,ttm,fs,fs0,fs1,fs2,fs3,fs4,fs5,fs6,fs7,fs8}. Cuprates are strongly correlated materials which shows abnormal properties in the underdoped and optimal doped regions of the phase diagram, above the superconducting transition temperature. The abnormal properties include T-linear resistivity at the optimal doping that possibly comes from strong electron-electron scattering. Similarly, there are many other abnormal properties like the temperature dependence of Hall coefficient, temperature dependent of NMR relaxation rate $1/T_1$ etc. \cite{dagotto}. The theoretical understanding of the above properties is not complete yet \cite{feature}.\\\\
The issue of non equilibrium electron relaxation in metals is being studied from long past and early investigations were done by  M. I. KAGANOV et.al. \cite{kaganov} . The major result of these investigations are that the rate of energy transfer from hotter electrons to colder lattice vibrations (for example, when the non equilibrium is created by a fast moving particle in a metallic sample) is proportional to $T_e^5-T^5$ at lower temperatures (when $T_e,T<<T_0$) and is proportional to $T_e-T$ at higher temperature (when $T_e,T>>T_0$). Here $T_e$ and $T$ are electron and phonon temperature respectively and $T_0$ is Debye temperature for metals. The theoretical model used in the above investigation is that of free electrons interacting with phonon degree of freedom \cite{kaganov}. The rate of energy change in superconducting material (BCS superconductors) has also been described by Philip B. Allen \cite{allen}, where the gap(s-wave) is introduced in the electron density of states.\\\\ 
In cuprates, there is d-wave symmetry of the gap function. So, the quasiparticles spectra is strongly momentum dependent. In the nodal direction, required excitation energy is zero and largest excitation energy is required for the antinodal direction. Due to zero nodal gap, nodal quasiparticles exist at any finite temperature $T$. Thus electron relaxation in cuprates is different from conventional BCS superconductors. In the present investigation we extend the work of M. I. KAGANOV et.al. and Allen's for the case of cuprates. \\\\
In literature, the behaviour of electron-phonon relaxation time has been studied for the case of metals. For that, it has been shown that the relaxation time follows the $T^{-3}$ law \cite{kaganov} ($T$ is phonon temperature) at low temperature ($T_e,T<<T_0$). The same evidence for relaxation time has also been reported experimentally for cuprates in low temperature region and it varies as $T^{-(3\pm0.5)}$ \cite{demsar,Jdemsar}.\\\\
In present work, we have studied the dynamics of relaxation process for cuprates in the pseudogap state. In section 2, the mathematical description of energy transfer from electrons to phonons have been discussed for both metals and cuprates in different temperature regions. In section 3, the results obtained for relaxation time for cuprates has been compared and detailed analysis of our work has been discussed with reference to the experimental work. At last, conclusion has been presented in section 4. 
\section{Mathematical Description for electron-phonon Relaxation Time}
When a laser pulse passes through an electronic material, it disturbs it and the system goes into temporal non-equilibrium state. In non-equilibrium state, electrons form (non) Fermi-Dirac distribution\footnote{Which deviate from equilibrium distribution} in which through electron-electron interaction, the electron subsystem relaxes to hot Fermi Dirac distribution\footnote{FD distribution at higher temperature}. Subsequently, to bring the whole system in equilibrium, electrons transfer their energy to phonons through electron-phonon relaxation process. 
\subsection{Metals}
In metals, electrons effectively behave as free particles and follow the dispersion relation as $\epsilon_k=\frac{\hbar^2 k^2}{2m}$, where $k$ is wave vector for electron and $m$ is effective mass of electrons \cite{kaganov}. In equilibrium, these electrons are distributed according to Fermi-Dirac distribution:
\begin{equation}
n_k=\frac{1}{(e^{\beta_e(\epsilon_k-\epsilon_0)}+1)}\hspace{1mm} ,\hspace{20mm} \beta_e=\frac{1}{k_BT_e}\hspace{1mm},
\end{equation}
and phonons follow Bose Einstein distribution:
\begin{equation}
n_f=\frac{1}{(e^{\beta \hbar \omega_f}-1)}\hspace{1mm},\hspace{20mm} \beta=\frac{1}{k_BT}\hspace{1mm},
\end{equation}
where $k_B$ is the Boltzmann constant, $T_e$ and $T$ are electron and phonon temperature respectively, $\omega_f (= s f)$ is phonon frequency, $s$ is velocity of sound, $f$ is phonon wave vector, $\epsilon_0 = \left(\frac{3n_0}{8\pi}\right)^{2/3}\frac{(2\pi\hbar)^2}{2m} $ is Fermi energy of the electrons \cite{kaganov} and $n_0$ is conduction electron density.\\ 
After the photo-excitation, electrons form non-thermal (non Fermi-Dirac) distribution which relax via electron-electron interactions to a hot thermal (Fermi-Dirac) distribution in a time scale called thermalization time scale, $\tau_{TH}$. Subsequently, this hot thermal distribution of electrons relax with phonons via electron-phonon interactions \cite{ttm}. The time taken for this process is called electron-phonon relaxation time, $\tau_{ep}$. In metals, it is generally assumed that $\tau_{ep}>>\tau_{TH}$. In such a case electrons thermalize within themselves very quickly, and electron-phonon relaxation takes larger time. In this case, the energy transferred \cite{kaganov} by electrons to phonons per unit time per unit volume is:        
\begin{equation}
\dot{E} = \int \dot N_f \hbar \omega_f (2\pi)^{-3}\hspace{1mm} V d^3f.
\end{equation}
Here $\dot N_f$ is the rate of change of number of phonons per unit time per unit volume in the system. This change is due to absorption and emission of phonons in the system.
Due to the absorption process, the rate of change of phonons per unit time per unit volume is:
\begin{equation}
-W_{k,k^{'}} \delta(\epsilon_{k^{'}}-\epsilon_k - \hbar \omega_f) n_k(1-n_{k^{'}}) n_f, 
\end{equation}
and for the emission process: 
\begin{equation}
W_{k^{'},k} \delta(\epsilon_k-\epsilon_{k^{'}}+ \hbar \omega_f) n_{k^{'}}(1-n_k) (n_f+1), 
\end{equation} 
where $W_{k,k^{'}} $ is the  transition  probability per unit time of scattering of an electron from the state with wave vector $k$ to $k^{'}$. Due to microscopic reversibility one has $ W_{k,k^{'}}=W_{k^{'},k}\left(=\frac{\pi U_{ep}^2 \omega_f}{\rho V s^2} \right)$. Here $U_{ep}$ is electron-phonon interaction energy, $V$ is volume of crystal lattice, $\rho$ is density of the material \cite{kaganov}.
Hence from equation (4) and (5), the total change in number of phonons per unit time per unit volume comes out to be;
\begin{equation}
\dot N_f = \int2\hspace{2mm} W_{k,k^{'}} \left\lbrace (n_f + 1) n_{k^{'}} (1-n_k)-n_f n_k(1-n_{k^{'}}) \right\rbrace \delta(\epsilon_k - \epsilon_{k^{'}}+ \hbar \omega_f ) (2\pi)^{-3}  d^3k^{'}.  
\end{equation} 
This is called Bloch-Boltzmann-Peierls equation \cite{allen} which on further simplification becomes \cite{kaganov}:
\begin{equation}
\dot N_f = \frac{m^2 U_{ep}^2\hbar \omega_f}{2\pi \hbar^4 \rho V s}\left[\frac{1}{(e^{\beta_e \hbar \omega_f} -1)}-\frac{1}{(e^{\beta \hbar \omega_f} -1)} \right]. 
\end{equation}
Inserting $\dot N_f$ in equation (3) results:
\begin{equation}
\dot{ E} = \int \frac{m^2 U_{ep}^2\hbar \omega_f}{2\pi \hbar^4 \rho V s}\left[\frac{1}{(e^{\beta_e \hbar \omega_f} -1)}-\frac{1}{(e^{\beta \hbar \omega_f} -1)} \right]  \hbar \omega_f (2\pi)^{-3}\hspace{1mm}V d^3f.
\end{equation}
This equation can be simplified in two temperature regimes:\\
Case I: For low temperature, $T,T_e<<T_0$ ($T_0$ is Debye temperature) 
\begin{equation}
\dot{E}=\frac{2I_2m^2 U_{ep}^2 (k_BT_0)^5}{(2\pi)^3\hbar^7\rho s^4}\left(\frac{T_e^5-T^5}{T_0^5} \right). 
\end{equation}\\
Case II: For high temperature, $T,T_e>>T_0$
\begin{equation}
\dot{E}=\frac{m^2 U_{ep}^2 (k_BT_0)^5}{2(2\pi)^3\hbar^7\rho s^4}\left(\frac{T_e-T}{T_0} \right). 
\end{equation}
Using above energy transfer relations, it is easy to calculate the relaxation time under near equilibrium condition $T_e - T<<T$: 
\begin{equation}
\tau_{ep}=\alpha \frac{8\pi^5 \hbar^7 \rho s^4 n_0}{15m^2U_{ep}^2\epsilon_0k^3}\frac{1}{T^3}.
\end{equation}
Here $\alpha=3/8$ (for low temperature) and $\alpha=3/2$ (for high temperature).
This relation reveals that the relaxation time varies as a function of temperature as $T^{-3}$ \cite{kaganov}. In next subsection we investigate this behavior for the case of cuprates. 
\subsection{Cuprates}
In cuprates, the dynamics for the relaxation process is quite different from metals. Here due to the presence of pseudogap, the nodal quasiparticles  at $(\pm\pi,\pm\pi)$ position of Brillouin zone follows different electronic dispersion. The nodal quasiparticles follow the linear dispersion relation as $\epsilon_k=\hbar v_F \vert k \vert$, where $v_F$ is the Fermi velocity for cuprates \cite{dispersion}. With this dispersion relation and under the condition $k^{'}\geq f_0/2$, in equation (6), the integral over $\theta$ can be simplified as (Detailed calculation is given in appendix A) :
\begin{equation}
\int_0^\pi\delta(\epsilon_k-\epsilon_{k^{'}}+ \hbar \omega_f)\sin\theta d\theta = \frac{1}{\hbar v_F f}.
\end{equation}
With this integral, equation (6) becomes
\begin{equation}
\dot N_f =\frac{2}{(2\pi)^{2}} \int_{\frac{f_{0}}{2}}^{\infty}\hspace{2mm} W_{k,k^{'}} \left\lbrace (n_f + 1) n_{k^{'}} (1-n_k)-n_f n_k(1-n_{k^{'}}) \right\rbrace\hspace{1mm} k^{'2}\hspace{1mm}dk^{'} \frac{1}{\hbar v_F f}.  
\end{equation}
Here $\epsilon_k$ is determined by energy conservation relation $\epsilon_{k^{'}}=\epsilon_k+\hbar\omega_f$. Now, using equation (1) and (2), equation(13) can be rewritten as:
\begin{equation}
\dot N_f =\frac{2}{(2\pi)^{2}}W_{k,k^{'}}\frac{1}{\hbar v_F f}\left[\frac{e^{\beta \hbar \omega_f} - e^{\beta_e \hbar \omega_f}}{e^{\beta \hbar \omega_f} -1} \right] \int_{\frac{f_{0}}{2}}^{\infty} \frac{e^{\beta_e (\epsilon_{k^{'}} - \epsilon_0 - \hbar \omega_f)}}{(e^{\beta_e (\epsilon_{k^{'}} - \epsilon_0 - \hbar \omega_f)}+1)(e^{\beta_e (\epsilon_{k^{'}} - \epsilon_0)}+1)}k^{'2} dk^{'}.
\end{equation}     
As electron energy ($\epsilon_{k^{'}} - \epsilon_0$) is much greater than the phonon energy $\hbar \omega_f$, this equation takes form (Details are in appendix A):
\begin{eqnarray}
\dot N_f &=& \frac{3U_{ep}^2 \omega_f^3}{2\pi \hbar v_F^4 \rho V s}\left[ \frac{1}{(e^{\beta_e \hbar \omega_f} -1)^3} - \frac{1}{(e^{\beta_e \hbar \omega_f} -1)^2 (e^{\beta \hbar \omega_f} -1)}  \right]\nonumber\\
&+&\frac{U_{ep}^2\epsilon_0^2}{2\pi \hbar^3 v_F^4 \rho V s} \omega_f \left[\frac{1}{(e^{\beta_e \hbar \omega_f} -1)}-\frac{1}{(e^{\beta \hbar \omega_f} -1)} \right].
\end{eqnarray}
Inserting the above expression for $\dot{N_f}$ into equation (3) one obtains $\dot{E}=\dot{E}_1$ + $\dot{E}_2$.\\
\begin{eqnarray}
\hspace{5mm}\dot{E}_1 &=& \int_0^{\omega_0} \frac{3 U_{ep}^2 \omega_f^3}{2\pi \hbar v_F^4 \rho V s}\left[ \frac{1}{(e^{\beta_e \hbar \omega_f} -1)^3} - \frac{1}{(e^{\beta_e \hbar \omega_f} -1)^2 (e^{\beta \hbar \omega_f} -1)}\right] \hbar \omega_f (2\pi)^{-3} V d^3f,\nonumber\\\nonumber\\
\hspace{10mm}\dot{E}_2 &=& \int_0^{\omega_0}\frac{U_{ep}^2\epsilon_0^2}{2\pi \hbar^3 v_F^4 \rho V s} \omega_f \left[\frac{1}{(e^{\beta_e \hbar \omega_f} -1)}-\frac{1}{(e^{\beta \hbar \omega_f} -1)} \right]\hbar \omega_f (2\pi)^{-3} V d^3f.\nonumber
\end{eqnarray}
Using $\omega_f = sf$ and $\beta_e \hbar \omega \approx e^{\beta_e \hbar \omega}-1$, $\dot{E}_1$ and $\dot{E}_2$ can be expressed as:\\
\begin{equation}
\hspace{10mm}\dot{E}_1 = \frac{6U_{ep}^2}{(2\pi)^3 v_F^4 \rho s^4}\int_0^{\omega_0}\left[ \frac{\omega_f^6 }{(e^{\beta_e \hbar \omega_f} -1)^3} - \frac{1}{(\beta_e \hbar)^2}.\frac{\omega_f^4 }{(e^{\beta \hbar \omega_f} -1)} \right]d\omega_f,  
\end{equation}\\
\begin{equation}
\text{and}\hspace{5mm}\dot{E_2} = \frac{2 U_{ep}^2 \epsilon_0^2}{(2\pi)^3 \hbar^2 v_F^4\rho s^4} \int_0^{\omega_0} \omega_f^4 \left[\frac{1}{(e^{\beta_e \hbar \omega_f} -1)}-\frac{1}{(e^{\beta \hbar \omega_f} -1)} \right]d\omega_f. 
\end{equation} 
Now, these equations can be solved in terms of Debye temperature $T_0$ in two different temperature regimes as discussed earlier in the case of metals.\\\\
Case I: For low temperature  $T,T_e << T_0$,
\begin{eqnarray}
\dot{E}_1 &=& \frac{6 U_{ep}^2(k_B T_0)^7}{(2\pi)^3 \hbar^7 v_F^4 \rho s^4}\left(\frac{T_e^7I_1-T^5T_e^2I_2}{T_0^7} \right),\\\nonumber\\  
\dot{E}_2 &=& \frac{2 U_{ep}^2 \epsilon_0^2(k_B T_0)^5}{(2\pi)^3 \hbar^7 v_F^4 \rho s^4}\left(\frac{T_e^5-T^5}{T_0^5} \right)I_2.  
\end{eqnarray}
Case II: For high temperature $T,T_e >> T_0$,
\begin{eqnarray}
\dot{E}_1 &=& \frac{3 U_{ep}^2(k_B T_0)^7}{2(2\pi)^3 \hbar^7 v_F^4 \rho s^4}\left(\frac{T_e^3-TT_e^2}{T_0^3} \right),\\\nonumber\\  
\dot{E}_2 &=& \frac{U_{ep}^2 \epsilon_0^2(k_B T_0)^5}{2(2\pi)^3 \hbar^7 v_F^4 \rho s^4}\left(\frac{T_e-T}{T_0} \right).  
\end{eqnarray}
Here $I_1$ and $I_2$ are definite integrals 
\begin{equation}
I_1 = \int_0^\infty \frac{x^6}{(e^x - 1)^3} dx,\hspace{10mm}I_2 = \int_0^\infty \frac{x^4}{e^x - 1} dx.\nonumber
\end{equation} 
These expression reveals that at low temperature the rate of energy transfer from electrons to phonons follows the temperature as $T^5$ and $T^7$, while at higher temperature it follows as $T$ and $T^3$.
\section{Comparison of Theoretical and Experimental Results in the case of cuprates}
In this section, we compare our results for electron-phonon relaxation time ($\tau_{ep}$) with experiment \cite{Jdemsar,demsar}. First let us discuss the behavior of energy transferred per unit volume per second from electrons to phonons $(\dot{E})$ at low and high temperature regimes for cuprates.\\\\
For numerical computation we use the following physical parameters for cuprates.  
Electron-phonon interaction energy $U_{ep}=50$ meV; Fermi energy for cuprates $\epsilon_0=0.2$ eV; Density for cuprates $\rho=6.2\times 10^3$ kg/m$^3$; Fermi velocity for cuprates $v_F=8\times 10^4$ m/s \cite{saxena}; Electron number density for cuprates $n_0=6.2\times 10^{27}$ /m$^3$; Debye temperature for cuprates $T_0=300$ K \cite{Shukor}. With these parameters $\dot{E}_1$ and  $\dot{E}_2$ are plotted in figure (3) (in Appendix B).  We can see that the magnitude of $\dot{E}_1$($\sim 10^{14}$ J/m$^3$ sec.) is very small compared to $\dot{E}_2$($\sim 10^{18}$ J/m$^3$ sec.), so the contribution due to $\dot{E}_1$ is negligible. Similarly, $\dot{E}_2$ term in high temperature region dominates as has been shown in figure (4) of Appendix B. Thus in the subsequent calculation we will use $\dot{E}_2$ for low and high  temperature regimes.\\
Hence the total rate of energy transferred from electrons to phonons is:\\\\
For low temperature:\\
\begin{equation}
\dot{E} = \frac{2 U_{ep}^2 \epsilon_0^2(k_B T_0)^5}{(2\pi)^3 \hbar^7 v_F^4 \rho s^4}\left(\frac{T_e^5-T^5}{T_0^5} \right)I_2.  
\end{equation} \\
For high temperature:\\
\begin{equation}
\dot{E} = \frac{U_{ep}^2 \epsilon_0^2(k_B T_0)^5}{2(2\pi)^3 \hbar^7 v_F^4 \rho s^4}\left(\frac{T_e-T}{T_0} \right).
\end{equation}
These expressions show that at low temperature energy transfer rate from electrons to phonons  
follows the temperature as $T^5$, while at high temperature it follows as proportional to $T$.
This behavior is equivalent to that in metals and it corroborates experimental findings \cite{Jdemsar,demsar} as discussed below.
\subsection{Relaxation Time for cuprates}
When a femtosecond pump pulse passes through the sample, the rate of change of energy in the electronic subsystem and the energy transfer rate to phonons \cite{kaganov} can be written as:  
\begin{equation}
C_e\dot{T}_e=-\dot{E}.
\end{equation}
This depicts the steady state condition. In steady state, amount of energy lost by electrons in a given interval of time is equal to the amount of energy gained by phonons.
From here one can solve $T_e$ as a function of time ($\dot{T}_e$ is rate of change of electron temperature and $C_e\left(=(\frac{\pi}{2})^2k_B^2n_0T_e/\epsilon_0\right)$  is electronic specific heat).
Since, we are interested in low temperature regime, as in the experimental paper \cite{Jdemsar,demsar} equation (24) ( by using of equation (22)) can be written as:
\begin{eqnarray}
C_e\frac{d{T}_e}{dt}= -\frac{2 U_{ep}^2 \epsilon_0^2(k_B T_0)^5}{(2\pi)^3 \hbar^7 v_F^4 \rho s^4}\left(\frac{T_e^5-T^5}{T_0^5} \right)I_2. 
\end{eqnarray}
For near equilibrium condition $T_e - T << T$, the relaxation time can be evaluated. 
Let us take $T_e - T=y$ and $A=\frac{2 U_{ep}^2 \epsilon_0^2(k_B T_0)^5}{(2\pi)^3 \hbar^7 v_F^4 \rho s^4}I_1$, then above equation will be simplified as:
\begin{equation}
\frac{dy}{dt}=-\frac{A}{C_eT_0^5}\left((T+y)^5 -T^5 \right).\nonumber 
\end{equation}
As $y<<T$, then expanding the above equation we have,
\begin{equation}
\frac{dy}{dt}=-\frac{5A}{C_eT_0^5}T^4y.
\end{equation}
This is a first order differential equation which gives solution:
\begin{equation}
y=T_e-T=ke^{-t/\tau_{ep}},
\end{equation}
where $k$ is integral constant and $\tau_{ep}$ is electro-phonon relaxation time which is a function of temperature and can be written as:
\begin{equation}
\tau_{ep}=\frac{\pi^5}{5}\frac{\hbar^7v_F^4\rho s^4 n_0}{\epsilon_0^3 U_{ep}^2 k_B^3I_2}\frac{1}{T^3}.
\end{equation}
This shows that electron-phonon relaxation time varies as $T^{-3}$ with the temperature. A numerical solution of equation (25) with the realistic initial conditions from the experiment \cite{demsar,Jdemsar} is presented in Appendix C and it agrees well with equation (27).
Experimentally, Schneider et.al \cite{Jdemsar,demsar} has discussed the relaxation time $\tau_{ep}$ using ultrafast femtosecond laser technique. They have taken cuprate samples at different doping and plotted electron-phonon relaxation time at low temperature regimes. They have described that at low temperature relaxation time varies as $T^{-3\pm0.5}$ \cite{Jdemsar,demsar} for both superconductor and non-superconductor samples (refer to figure (1)). From our theoretical calculation we have shown that the electron-phonon relaxation time follows as $T^{-3}$ for cuprates. Thus experimental findings of reference  \cite{Jdemsar,demsar} are corroborated by the theory presented here (refer to figure (1)).
For the best fit we have used $U_{ep}=50$ meV and this value of electron-phonon interaction energy agrees with that obtained from inelastic neutron scattering experiment \cite{Roch}
\begin{figure}[htbp]
	\centering
	\includegraphics[scale=0.5]{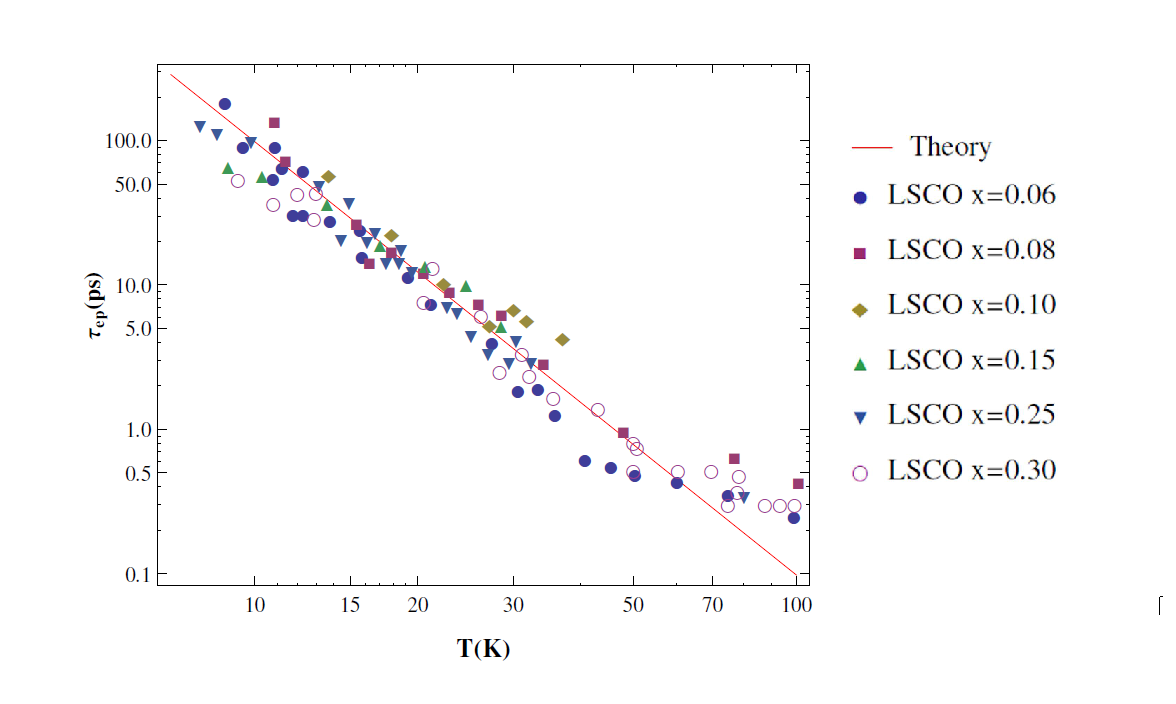}
	\caption{Relaxation time $\tau_{ep}$ for LSCO at different dopings (symbols from experiment) and its comparison with present Theory (solid line). }
\end{figure}
\newpage
\section{Conclusion}
In reference \cite{Jdemsar,demsar}, authors conclude from their experimental results that $\tau_{ep}\propto T^{-3\pm 0.5}$. They observe that this divergence sets in the normal state of the superconducting samples. And this continue to the superconducting state. The authors of the experimental paper \cite{Jdemsar,demsar} also find that the divergence $\tau_{ep}\propto T^{-3\pm 0.5}$ holds both for single layer cuprates La$_{2-x}$Sr$_x$CuO$_4$, Bi$_2$Sr$_2$CuO$_{6+z}$  and double layer cuprates Bi$_2$Sr$_2$CaCu$_2$O$_{8+\delta}$. Thus, this points to the universality of $\tau_{ep}\propto T^{-3\pm 0.5}$.\\\\
In the present theoretical investigation, we have reproduced this divergence, namely $\tau_{ep}\propto T^{-3}$. The microscopic model used was that of hot Fermi-Dirac distribution of free electrons relaxing through electron-phonon interactions. The electronic dispersion used is the linear one $\epsilon_k=\hbar v_F \vert k\vert$ \cite{dispersion}. Within this microscopic setting and using Bloch-Boltzmann-Peierls kinetic equation we are able to reproduce the experimental results. For quantitative fitting we varied the electron-phonon interaction energy $U_{ep}$ and best fit results are obtained for $U_{ep}$=50 meV. This magnitude roughly agrees with inelastic neutron scattering experiments \cite{Roch}. \\ \\\\  
{\bf Acknowledgment}\\
My thanks to Dr. Nabyendu Das for reading the manuscript and discussion.

\pagebreak 
\section*{Appendix A}
The energy transferred by electrons to phonons per unit time per unit volume is:        
\begin{equation}
\dot{E} = \int \dot N_f \hbar \omega_f (2\pi)^{-3}\hspace{1mm}V d^3f.
\end{equation}
Here $\dot N_f$ is total change in number of phonons per unit time per unit volume in the system which is:
\begin{equation}
\dot N_f = \int2\hspace{2mm} W_{k,k^{'}} \left\lbrace (n_f + 1) n_{k^{'}} (1-n_k)-n_f n_k(1-n_{k^{'}}) \right\rbrace \delta(\epsilon_k - \epsilon_{k^{'}}+ \hbar \omega_f ) (2\pi)^{-3}  d^3k^{'}.  
\end{equation} 
For calculation of delta integral we use energy and momentum conservation which are:
\begin{eqnarray}
\epsilon_{k^{'}}&=&\epsilon_k + \hbar \omega_f,\nonumber\\
\vec{k}^{'}&=& \vec{k} + \vec{f}.\nonumber
\end{eqnarray} 
\begin{figure}[htbp]
	\centering
	\includegraphics[scale=0.3]{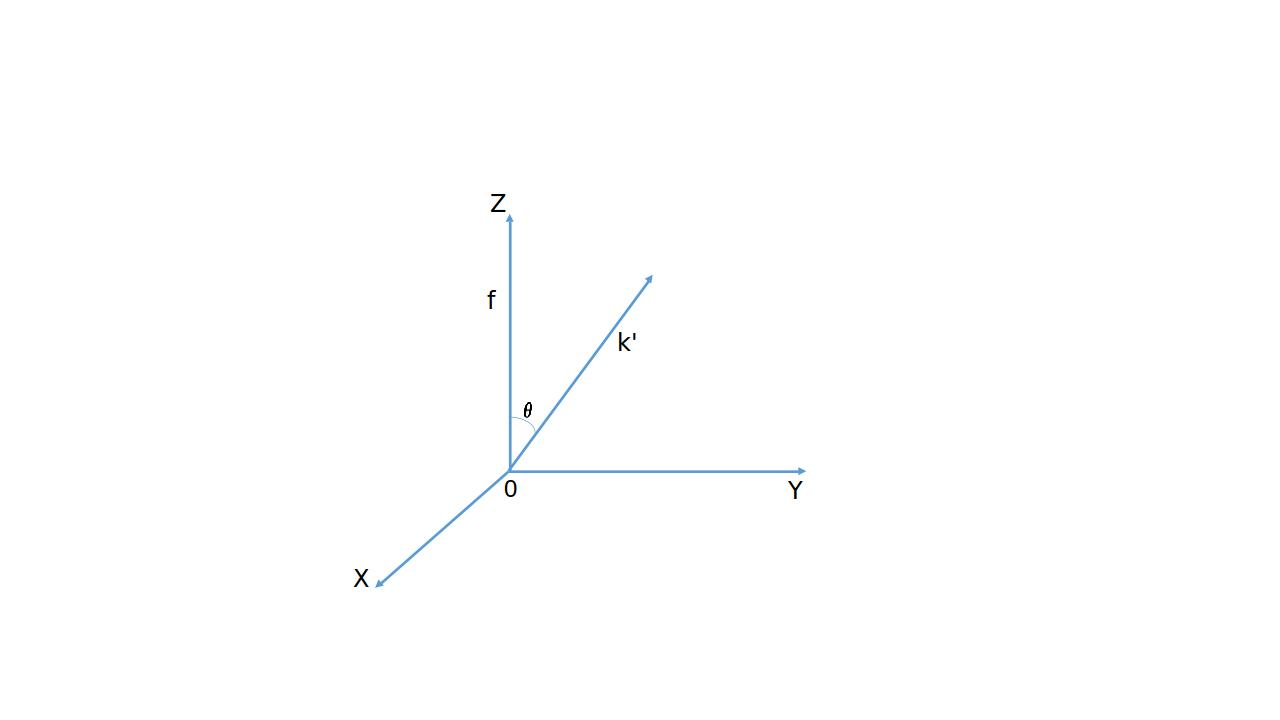}
	\caption{Phonon wave vector is in z direction and angle $\theta$ is between electron wave vector ($k^{'}$) and phonon wave vector($f$).}
\end{figure}
If scattering angle between $\vec{k}^{'}$ and $\vec{f}$ is $\theta$ we can write
\begin{equation}
k^2 = k^{'2} + f^2 - 2 k^{'}f \cos\theta.
\end{equation}
Here $k$ and $f$ are magnitudes of electron and phonon wave vector respectively. Now from energy conservation 
\begin{eqnarray}
\epsilon_{k^{'}}-\epsilon_k - \hbar \omega_f &=& 0,\nonumber\\
\hbar v_F (k^{'}- k) - \hbar \omega_f &=& 0,
\end{eqnarray}
where dispersion relation for cuprates is
\begin{equation}
\epsilon_k = \hbar v_F \vert k \vert.\nonumber
\end{equation}
Now by the use of equation (31) in equation (32) we will get:
\begin{eqnarray}
\hbar v_F \left( k^{'}-\sqrt{k^{'2} + f^2 - 2 k^{'}f \cos\theta}\right)  - \hbar \omega_f &=&0,\\
\left( k^{'}-\sqrt{k^{'2} + f^2 - 2 k^{'}f \cos\theta}\right)  - \frac{s}{v_F}f &=&0.
\end{eqnarray}
Here we have used $\omega_f = sf$ and $\frac{s}{v_F}\rightarrow0$ as $v_F>>s$ then we will get one condition for $\theta$
\begin{equation}
\cos\theta = \frac{f}{2k^{'}},
\end{equation}
\begin{equation}
0\leq\theta\leq\frac{\pi}{2}.
\end{equation}
So under this condition there will be a minimum wave vector for electrons in term of minimum wave vector  of phonon which can be written like this
\begin{eqnarray}
\frac{f}{2k^{'}} &\leq & 1.
\end{eqnarray}
Here
\begin{equation}
k^{'}_{min}\simeq\frac{f}{2},\hspace{2mm}\text{for}\hspace{2mm}\theta=0, 
\end{equation}
\begin{equation}
k^{'}_{max}\simeq\infty,\hspace{2mm} \text{for} \hspace{2mm}\theta=\frac{\pi}{2}.
\end{equation}
Thus, electrons that are having minimum wave vector $k_0=f_0/2$ ($f_0$ is minimum wave vector for phonons) and above this will take part in the interaction. Under this condition delta integral can be calculated as follows:
\begin{eqnarray}
\int_0^\pi\delta(\epsilon_k-\epsilon_{k^{'}}+ \hbar \omega_f)\sin\theta d\theta &=&\int_0^\pi\delta\left(\hbar v_F \left( k^{'}-\sqrt{k^{'2} + f^2 - 2 k^{'}f \cos\theta}\right)  - \hbar s f \right) \sin\theta d\theta,\nonumber\\
\int_0^\pi\delta(\epsilon_k-\epsilon_{k^{'}}+ \hbar \omega_f)\sin\theta d\theta &\simeq&\frac{1}{\hbar v_F k^{'}}\int_0^\pi\delta\left(1-\sqrt{1 + \left(\frac{f}{k^{'}}\right)^2 - 2\frac{f}{k^{'}} \cos\theta}\right)\sin\theta d\theta.
\end{eqnarray}
Here we have also used $\frac{s}{v_F}\rightarrow0$ as $v_F>>s$. Define
\begin{equation}
t=1-\sqrt{1 + \left(\frac{f}{k^{'}}\right)^2 - 2\frac{f}{k^{'}}\cos\theta}\nonumber.
\end{equation}
Now in terms of $t$ equation (40) can be written as
\begin{equation}
\int_0^\pi\delta(\epsilon_k-\epsilon_{k^{'}}+ \hbar \omega_f)\sin\theta d\theta \simeq\frac{1}{\hbar v_F f}\int_{-\frac{f}{k^{'}}}^{\frac{f}{k^{'}}}\delta(t)(1-t)dt\nonumber.
\end{equation} 
So this delta integral becomes:
\begin{equation}
\int_0^\pi\delta(\epsilon_k-\epsilon_{k^{'}}+ \hbar \omega_f)\sin\theta d\theta \simeq\frac{1}{\hbar v_F f}.
\end{equation} 
Therefore the total change in number of phonons per unit time per unit volume from equation (30) will be:
\begin{equation}
\dot N_f =\frac{2}{(2\pi)^{2}} \int_{\frac{f_{0}}{2}}^{\infty}\hspace{2mm} W_{k,k^{'}} \left\lbrace (n_f + 1) n_{k^{'}} (1-n_k)-n_f n_k(1-n_{k^{'}}) \right\rbrace\hspace{1mm} k^{'2}\hspace{1mm}dk^{'} \frac{1}{\hbar v_F f}.  
\end{equation}
If phonon wave vector $\vec{f}$ and the initial electron wave vector $\vec{k^{'}}$ are given then final wave vector $\vec{k}$ is fixed due to momentum conservation condition. Thus, $\vec{k}$ in the above equation can be written in terms of $\vec{f}$ and $\vec{k^{'}}$.   
After inserting value of $n_k$ and $n_f$ this equation will be as follows:
\begin{equation}
\dot N_f =\frac{2}{(2\pi)^{2}}W_{k^{'}-f,k^{'}}\frac{1}{\hbar v_F f}\left[\frac{e^{\beta \hbar \omega_f} - e^{\beta_e \hbar \omega_f}}{e^{\beta \hbar \omega_f} -1} \right] \int_{\frac{f_{0}}{2}}^{\infty} \frac{e^{\beta_e (\epsilon_{k^{'}} - \epsilon_0 - \hbar \omega_f)}}{(e^{\beta_e (\epsilon_{k^{'}} - \epsilon_0 - \hbar \omega_f)}+1)(e^{\beta_e (\epsilon_{k^{'}} - \epsilon_0)}+1)}k^{'2} dk^{'}.
\end{equation}
Now to solve this integral let us take
\begin{eqnarray}
\beta_e (\epsilon_{k^{'}} - \epsilon_0) &=& u,\nonumber\\
\beta_e \hbar \omega_f &=& v,\nonumber\\
\epsilon_{k^{'}}&=& \hbar v_F \vert k^{'}\vert\nonumber.
\end{eqnarray}
Now in terms of $u$  equation (43) can be rewritten as:
\begin{equation}
\dot N_f = \frac{2}{(2\pi)^{2}}W_{k^{'}-f,k^{'}}\frac{1}{\hbar v_F f}\left[\frac{e^{\beta \hbar \omega_f} - e^{\beta_e \hbar \omega_f}}{e^{\beta \hbar \omega_f} -1} \right] \frac{1}{(\beta_e \hbar v_F)^3} \int_{u(f_0/2)}^{\infty} \frac{e^u}{(e^u+1)^2} (u+\beta_e \epsilon_0)^2 du.
\end{equation}
Where one important point is that $u>>v$ (electron energy is greater than the phonons). So, under this condition  $v \rightarrow 0$ and $u(f_0/2)\rightarrow -\infty$. In the integrand of equation (44), there will be three terms. In this second integral is odd function of $u$, so that first and third integral will contribute in this equation. Therefore change in  number of phonon in the system can be written as follows:
\begin{eqnarray}
\dot N_f &=& \frac{3 U_{ep}^2 \omega_f^3}{2\pi \hbar v_F^4 \rho V s}\left[ \frac{1}{(e^{\beta_e \hbar \omega_f} -1)^3} - \frac{1}{(e^{\beta_e \hbar \omega_f} -1)^2 (e^{\beta \hbar \omega_f} -1)}  \right]\nonumber\\
&+&\frac{U_{ep}^2\epsilon_0^2}{2\pi \hbar^3 v_F^4 \rho V s} \omega_f \left[\frac{1}{(e^{\beta_e \hbar \omega_f} -1)}-\frac{1}{(e^{\beta \hbar \omega_f} -1)} \right].
\end{eqnarray}
According to equation (29) the energy transferred $\dot{E}$ can be written as sum of two integral terms namely $\dot{E}_1$ and $\dot{E}_2$\\
\begin{eqnarray}
\dot{E}_1 &=& \int_0^{\omega_0} \frac{3 U_{ep}^2 \omega_f^3}{2\pi \hbar v_F^4 \rho V s}\left[ \frac{1}{(e^{\beta_e \hbar \omega_f} -1)^3} - \frac{1}{(e^{\beta_e \hbar \omega_f} -1)^2 (e^{\beta \hbar \omega_f} -1)}\right] \hbar \omega_f (2\pi)^{-3} V d^3f,\nonumber\\\nonumber\\
\dot{E}_2 &=& \int_0^{\omega_0}\frac{U_{ep}^2\epsilon_0^2}{2\pi \hbar^3 v_F^4 \rho V s} \omega_f \left[\frac{1}{(e^{\beta_e \hbar \omega_f} -1)}-\frac{1}{(e^{\beta \hbar \omega_f} -1)} \right]\hbar \omega_f (2\pi)^{-3} V d^3f.\nonumber
\end{eqnarray}\\\\
Taking $\omega_f = sf$ and $\beta_e \hbar \omega \approx (e^{\beta_e \hbar \omega}-1)$, energies $\dot {E_1}$ and $\dot {E_2}$ in terms of $\omega_f$ can be written as:\\
\begin{equation}
\dot{E}_1 = \frac{6U_{ep}^2}{(2\pi)^3 v_F^4 \rho s^4}\left[ \int_0^{\omega_0} \frac{\omega_f^6 d\omega_f}{(e^{\beta_e \hbar \omega_f} -1)^3} - \frac{1}{(\beta_e \hbar)^2}.\frac{\omega_f^4 d\omega_f}{(e^{\beta \hbar \omega_f} -1)} \right],  
\end{equation}
\begin{equation}
\dot E_2 = \frac{2 U_{ep}^2 \epsilon_0^2}{(2\pi)^3 \hbar^2 v_F^4\rho s^4} \int_0^{\omega_0} \omega_f^4 \left[\frac{1}{(e^{\beta_e \hbar \omega_f} -1)}-\frac{1}{(e^{\beta \hbar \omega_f} -1)} \right]d\omega_f. 
\end{equation}
Let us solve these equations in terms of Debye temperature $T_0$ in two different temperature regimes:\\\\
{\bf Case I: For low temperature when $T,T_e << T_0$,}\\\\
Here in the equation (46) we are taking $x=\beta_e\hbar\omega_f$ for first term and $x=\beta\hbar\omega_f$ for second term. $T_0=\frac{\hbar \omega_0}{k}$ is Debye temperature. Then this equation (46) will take form:
\begin{equation}
\dot{E}_1 = \frac{6U_{ep}^2(k_B T_0)^7}{(2\pi)^3 v_F^4 \rho s^4\hbar^7}\left[\left( \frac{T_e}{T_0}\right) ^7 \int_0^{T_0/T_e} \frac{x^6 dx}{(e^{x} -1)^3} - \left( \frac{T^5 T_e^2}{T_0^7}\right) \int_0^{T_0/T} \frac{x^4 dx}{(e^{x} -1)} \right].  
\end{equation}
Now simplification form of this equation will be:
\begin{eqnarray}
\dot{E}_1 &=& \frac{6 U_{ep}^2(k_B T_0)^7}{(2\pi)^3 \hbar^7 v_F^4 \rho s^4}\left(\frac{T_e^7I_1-T^5T_e^2I_2}{T_0^7} \right).\nonumber
\end{eqnarray}
Under the same condition mentioned above $\dot{E_2}$ can be simplified as:
\begin{eqnarray}  
\dot{E}_2 &=& \frac{2 U_{ep}^2 \epsilon_0^2(k_B T_0)^5}{(2\pi)^3 \hbar^7 v_F^4 \rho s^4}\left(\frac{T_e^5-T^5}{T_0^5} \right)I_2.\nonumber  
\end{eqnarray}
{\bf Case II: For high temperature when $T,T_e >> T_0$,}\\\\
Similar method can be used for high temperature but here we take one condition such that $e^x-1\approx x$, then transferred energy will be 
\begin{eqnarray}
\dot{E}_1 &=& \frac{3 U_{ep}^2(k_B T_0)^7}{2(2\pi)^3 \hbar^7 v_F^4 \rho s^4}\left(\frac{T_e^3-TT_e^2}{T_0^3} \right),\nonumber\\\nonumber\\  
\dot{E}_2 &=& \frac{U_{ep}^2 \epsilon_0^2(k_B T_0)^5}{2(2\pi)^3 \hbar^7 v_F^4 \rho s^4}\left(\frac{T_e-T}{T_0} \right).\nonumber  
\end{eqnarray}
Here $I_1$ and $I_2$ are definite integrals which are as follows
\begin{equation}
I_1 = \int_0^\infty \frac{x^6}{(e^x - 1)^3} dx,\hspace{10mm}I_2 = \int_0^\infty \frac{x^4}{e^x - 1} dx,\nonumber.
\end{equation}  
\pagebreak
\section*{Appendix B}
\begin{figure}[htbp]
	\centering
	\begin{subfigure}[htbp]{0.4\textwidth}
		\centering
		\includegraphics[width=\textwidth]{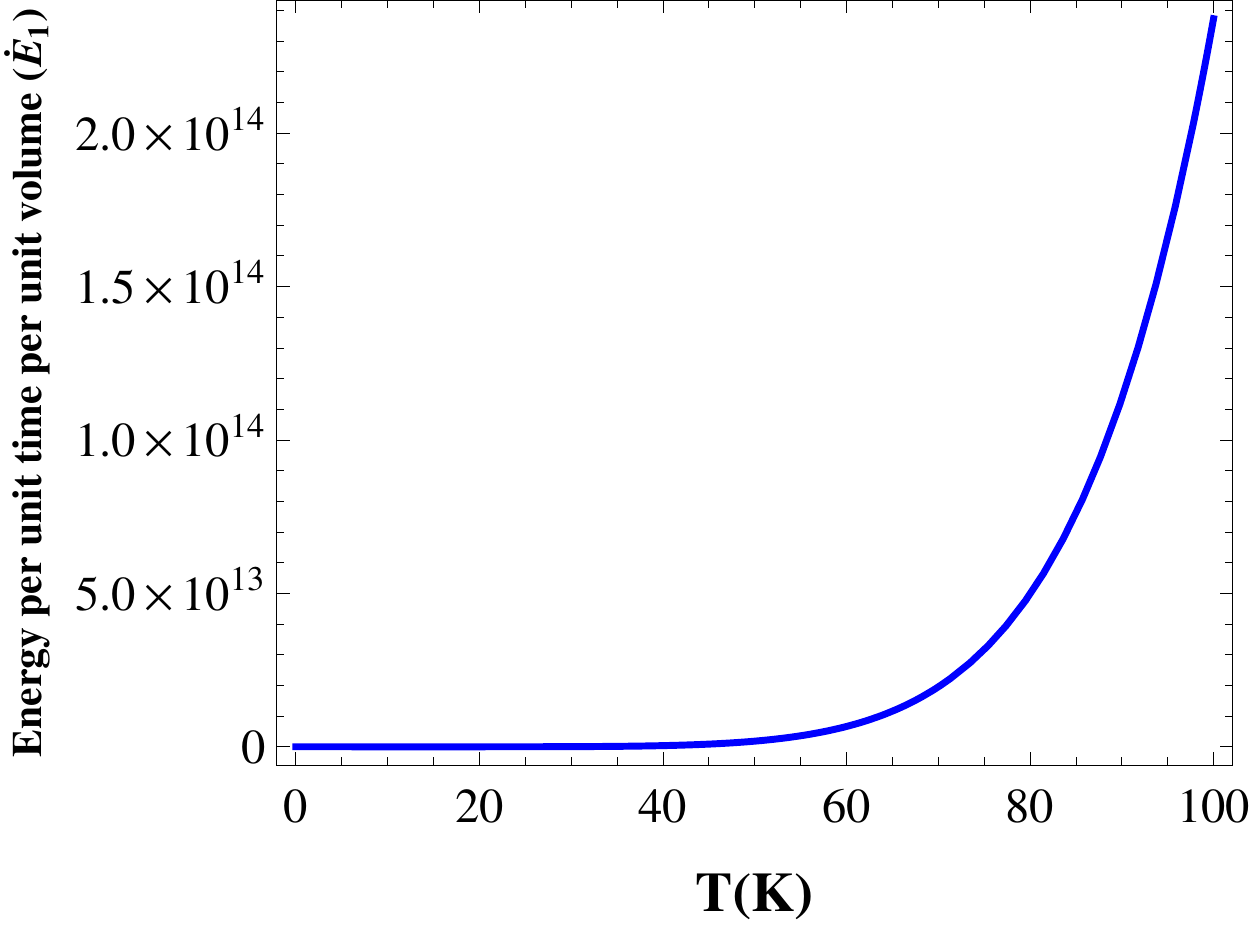}
		\caption{Variation of energy ($\dot{E}_1$) with T}
	\end{subfigure}
	\hfill
	\begin{subfigure}[htbp]{0.4\textwidth}
		\centering
		\includegraphics[width=\textwidth]{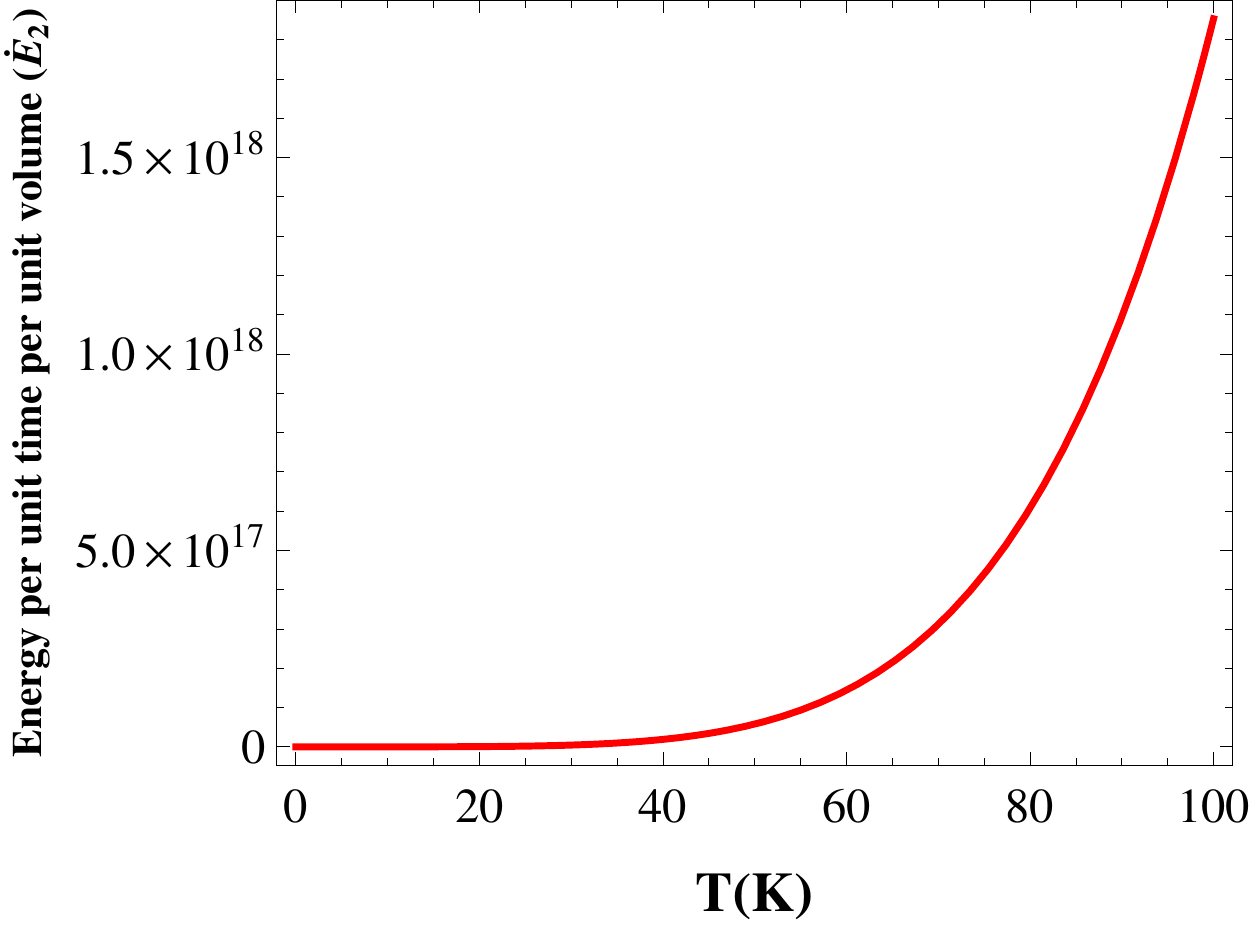}
		\caption{Variation of energy ($\dot{E}_2$) with T}
	\end{subfigure}
	\caption{Low temperature regime }
\end{figure}
	\hfill
\begin{figure}
	\centering
	\begin{subfigure}[htbp]{0.4\textwidth}
		\centering
		\includegraphics[width=\textwidth]{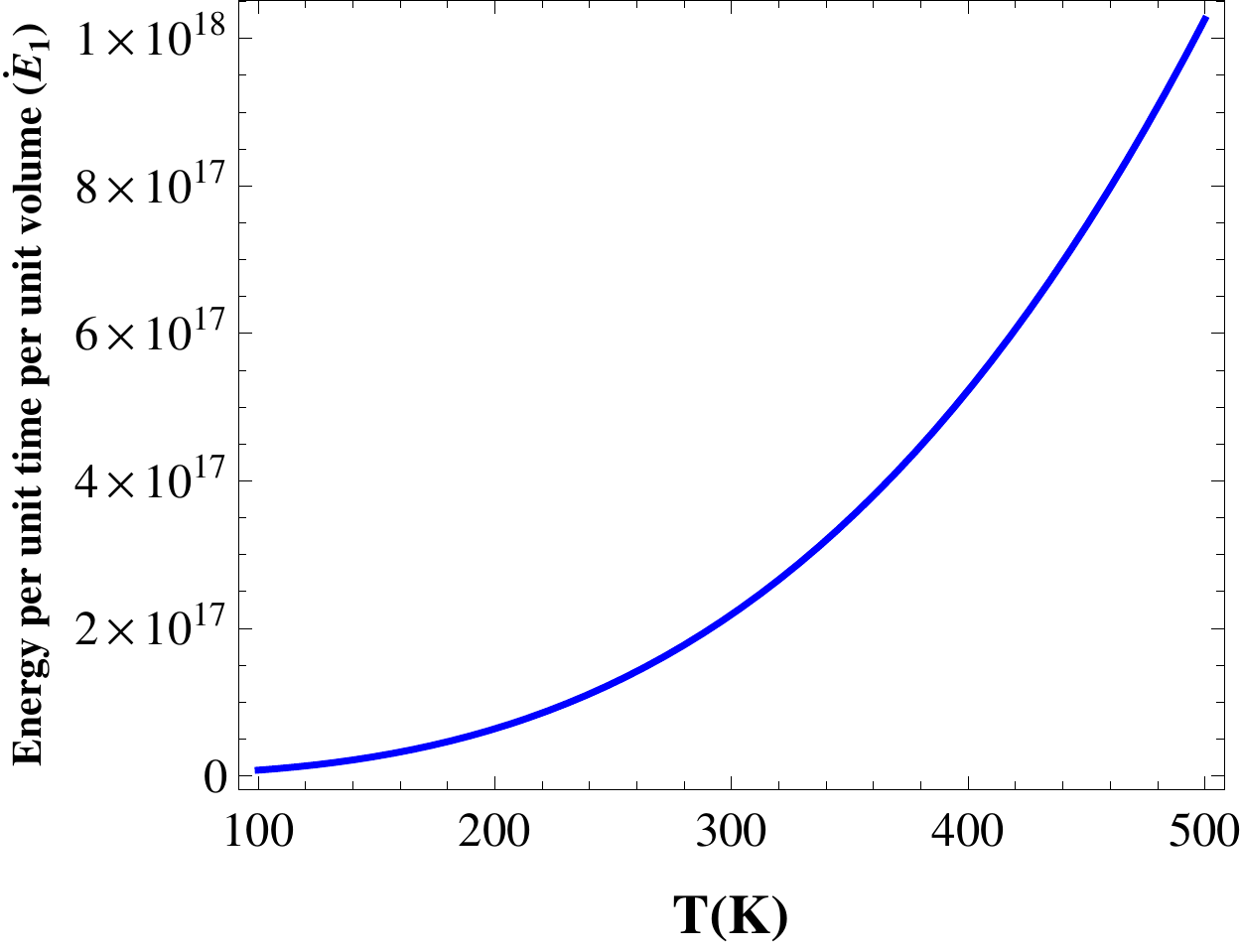}
		\caption{Variation of energy ($\dot{E}_1$) with T}
	\end{subfigure}
	\hfill
	\begin{subfigure}[htbp]{0.4\textwidth}
		\centering
		\includegraphics[width=\textwidth]{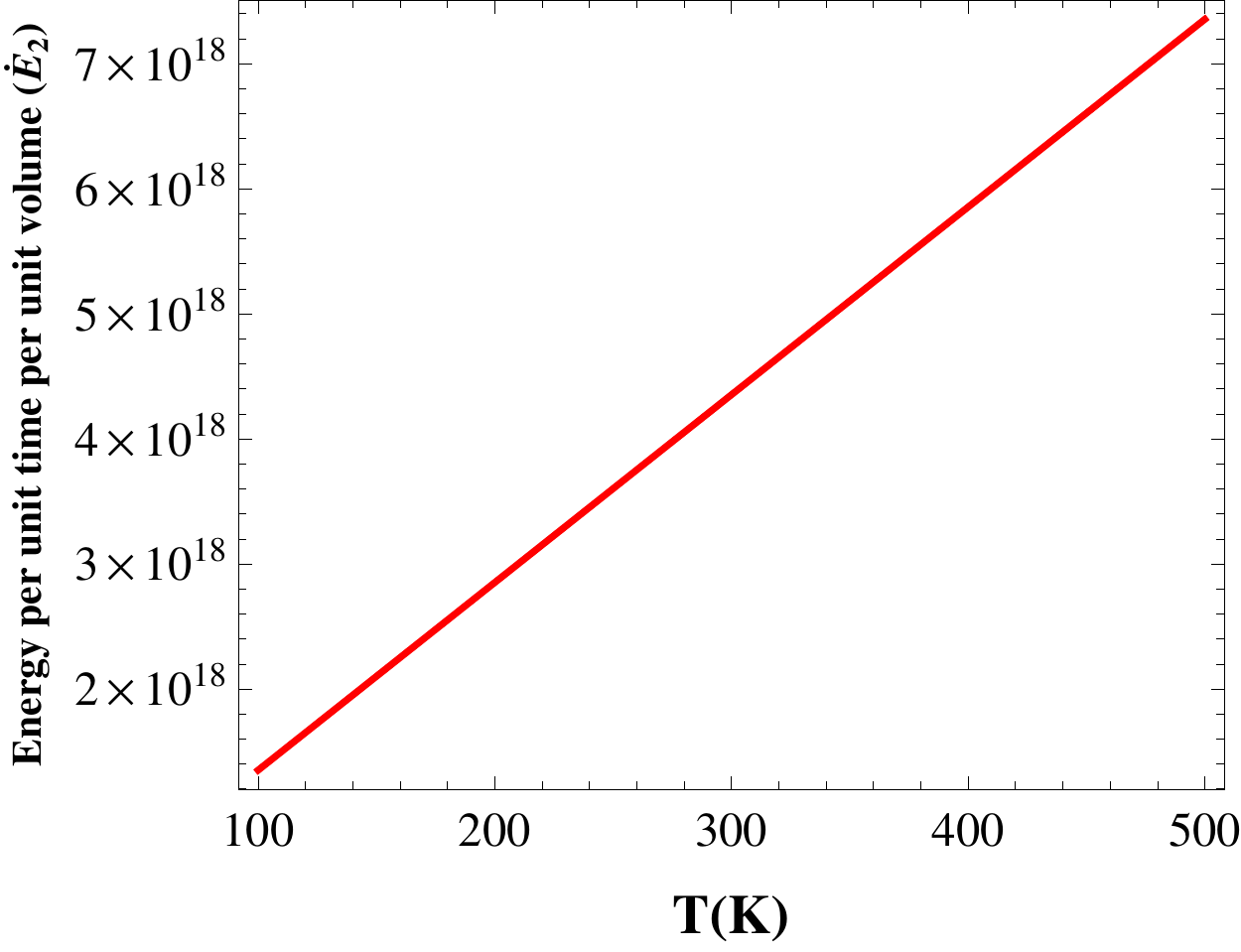}
		\caption{Variation of energy ($\dot{E}_2$) with T}
	\end{subfigure}
	\caption{High temperature regime}
\end{figure}
\pagebreak
\section*{Appendix C}
{\bf Electron temperature with time without taking near equilibrium condition}\\\\
In the experimental paper \cite{demsar,Jdemsar} they are taking pump pulse energy $5\times 10^{-12}$ joule and energy density $9\times 10^{-8}$ joule/cm$^2$. Sample absorbs $\sim 10\%$ of it. Width of the sample is taken to be 50nm (private communication with Prof.J. Demsar). So absorbed energy per unit volume by the sample will be equivalent to specific heat times temperature difference that can be written as:
\begin{equation}
\eta\frac{E}{V}=C_e(T_e-T)=\gamma T_e(T_e-T).
\end{equation}  
Here $\eta$ is absorption coefficient and $C_e=\gamma T_e=\frac{\pi^2k_B^2n_0}{4\epsilon_0}T_e$. With given data (in section 3), we can solve the above equation and it will be:
\begin{equation}
T_e^2 - TT_e - 19.79=0.
\end{equation}
This equation provides initial condition for equation (25) at different phonon  temperature $T$. With initial condition $T_e(t=0)$ equation (25)  
\begin{eqnarray}
C_e\frac{d{T}_e}{dt}= -\frac{2 U_{ep}^2 \epsilon_0^2(k_B T_0)^5}{(2\pi)^3 \hbar^7 v_F^4 \rho s^4}\left(\frac{T_e^5-T^5}{T_0^5} \right)I_2,\nonumber 
\end{eqnarray}
can be simplified to
\begin{equation}
\frac{d{T}_e}{dt}+2.03747\times 10^6\left(\frac{T_e^5-T^5}{T_e} \right)=0.
\end{equation}
A numerical solution of the above equation is presented in figure (5) for different phonon temperature $T$ (dotted lines). It is compared with approximate analytical solution for equation (27) in the same figure (5) (solid lines). For the calculation of $T_e(t)$ from equation (27), $\tau_{ep}$ is calculated from equation (28) for given $T$. Integration constant $k$ is determined from the initial condition $T_e(t=0)$.  
\begin{figure}[htbp]
	\centering
	\begin{subfigure}[htbp]{0.46\textwidth}
		\centering
		\includegraphics[width=\textwidth]{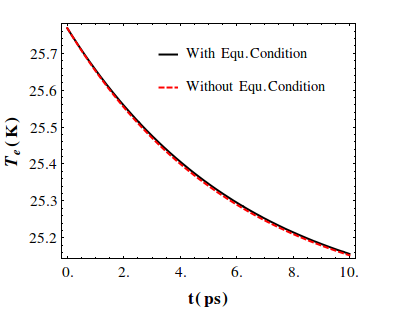}
		\caption{T=25 K}
	\end{subfigure}
	\hfill
	\begin{subfigure}[htbp]{0.46\textwidth}
		\centering
		\includegraphics[width=\textwidth]{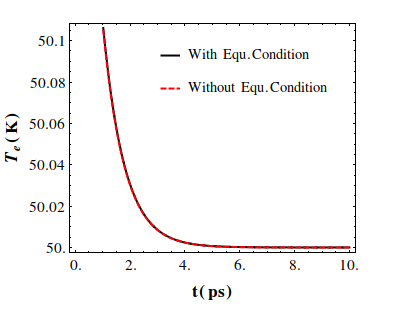}
		\caption{T=50 K}
	\end{subfigure}
	\hfill
	\begin{subfigure}[htbp]{0.46\textwidth}
		\centering
		\includegraphics[width=\textwidth]{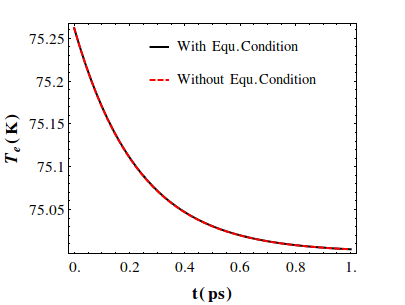}
		\caption{T=75 K}
	\end{subfigure}
	\hfill
	\begin{subfigure}[htbp]{0.46\textwidth}
		\centering
		\includegraphics[width=\textwidth]{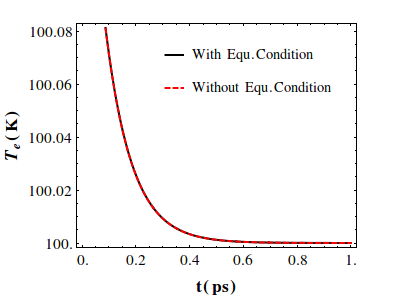}
		\caption{T=100 K}
	\end{subfigure}
	\caption{Time dependence of electron temperature both from numerical solution of equation (52) (dotted lines), and from approximate result of equation (27) (solid lines), where $\tau_{ep}$ calculated from equation (28) at different temperature $T$.}
\end{figure}                    
\end{document}